\documentstyle[12pt]{article}
\newcommand{\be}{\begin{equation}}
\newcommand{\ee}{\end{equation}}
 
\newcommand{\bea}{\begin{eqnarray}}
\newcommand{\eea}{\end{eqnarray}}
\newcommand{\ba}{\begin{array}}
\newcommand{\ea}{\end{array}}
\newcommand{\beqa}{\begin{eqnarray}}
\newcommand{\eeqa}{\end{eqnarray}}
 
\newcommand{\NP}[1]{Nucl. Phys.\ {\bf #1}\ }
\newcommand{\PL}[1]{Phys. Lett.\ {\bf #1}\ }

\newcommand{\PR}[1]{Phys. Rev.\ {\bf #1}\ }

\newcommand{\ZP}[1]{Z. Phys.\ {\bf #1}\ }

\newcommand{\Tr}{{\rm Tr}}

\newcommand{\D}{\delta}
\newcommand{\DE}{\Delta}

\newcommand{\tc}{\tilde\chi }
\newcommand{\tco}{\tilde\chi_1 }
\newcommand{\tct}{\tilde\chi_2 }
\newcommand{\lan}{\langle}
\newcommand{\ran}{\rangle}
\newcommand{\ssu}{$SU(2)_L\times SU(2)_R\times U(1)_{B-L}\,$}

\newcommand{\matr}{\left( \begin{array}}
\newcommand{\ematr}{\end{array} \right)}

\newcommand{\lsim}
{{\;\raise0.3ex\hbox{$<$\kern-0.75em\raise-1.1ex\hbox{$\sim$}}\;}}
\newcommand{\gsim}
{{\;\raise0.3ex\hbox{$>$\kern-0.75em\raise-1.1ex\hbox{$\sim$}}\;}}

\begin{document}

\begin{titlepage}

\begin{flushright}
\parbox[t]{4.8truecm} {\begin{center}  {\large  HIP-1997-24/TH}
\\
{(hep-ph/9705406)} 
\end{center}}
\end{flushright}
\vfill

\Large
 
\begin{center}
 {\bf Signals of
spontaneous R parity breaking at LEP and at a linear collider}
 
\bigskip
\normalsize {K. Huitu$^a$, J. Maalampi$^b$ and K. Puolam\"aki$^a$ }\\ [15pt] {\it
$^a$Helsinki Institute of Physics, P.O.Box 9, FIN-00014 University of Helsinki
\\$^b$Department of Physics, Theoretical Physics Division, P.O.Box 9,
FIN-00014 University of Helsinki}
 
{May 1997}

\vfill
 
\normalsize
 
{\bf\normalsize \bf Abstract}
 
 \end{center}

\normalsize

\noindent We study the production of neutralinos and charginos at LEP
and at a linear collider in the case of spontaneously broken
R-parity. We first investigate the constraints on the single
neutralino and chargino production from the LEP 1 experiments, and
then we consider the production at LEP 2 and at a linear collider.  We
concentrate on the supersymmetric $SU(2)_L \times SU(2)_R \times
U(1)_{B-L}$ model, where the spontaneous R-parity breaking is
inevitable and is associated with the breaking of the LR-symmetry.

\end{titlepage}
 
\newpage
 
\setcounter{page}{2}

\section{Introduction} 

The lepton and baryon number conservation of the Standard Model (SM)
is incorporated in the Minimal Supersymmetric Standard Model (MSSM) by
the conservation of R-parity, $R=(-1)^{3(B-L)+2S}$.  Under this
symmetry the standard model particles are R-even and their
superpartners R-odd. The gauge symmetry and supersymmetry do allow an
explicit R-parity breaking through three point interactions and
bilinear terms. These interactions would, however, violate either
lepton or baryon number.  If both are realized, the proton will decay
fast \cite{proton}. Therefore, in MSSM with broken R-parity it is
usually assumed that either the couplings that break the lepton number
or the couplings that break the baryon number vanish.

In addition to the above mentioned explicit breaking of R-parity, the
breaking can be spontaneous.  In MSSM the spontaneous breaking could
occur through a non-zero VEV of the scalar partner of a neutrino,
which would lead -- as the lepton number is not part of the gauge
symmetry -- to the emergence of a physical massless Goldstone boson, a
majoron. This would open a new invisible decay mode of the $Z$-boson
ruled out by LEP experiments \cite{LEPMajoron}. Phenomenology of
models where R-parity breaking is associated with an isosinglet
neutrino instead of a usual isodoublet neutrino and where the majoron
has a negligible coupling to the weak $Z$ boson, as well as models
with a new $U(1)$ symmetry with an extra $Z'$ absorbing the Goldstone
mode, have been studied in \cite{majoron,BBHH,AM}.

In the supersymmetric left-right model (SLRM) \cite{HMRF} with gauge
group $SU(2)_L\times SU(2)_R\times U(1)_{B-L}$, R-parity is conserved
at the level of Lagrangian \cite{autoR}. This is because all the terms
allowed by the $U(1)_{B-L}$ gauge symmetry are automatically
R-even. On the other hand, in the physically consistent minimum one of
the neutrinos necessarily has a VEV and R-parity is thus spontaneously
broken \cite{KM,HM}.  As the R-parity breaking arises in this model
through one or more of the sneutrinos getting a nonzero VEV, only the
lepton number is violated and the breaking has no effect on the proton
decay. The three point interactions of the MSSM with an explicit
R-parity violation are experimentally strictly constrained (see
e.g. \cite{CH}). In models with spontaneous R-parity violation, like
SLRM, these couplings are naturally suppressed since they are
proportional to small mixing angles or they are generated by loops.

An important consequence of the spontaneous breaking of the R-parity
is that R-odd particles, i.e. supersymmetric particles not present in
the Standard Model, can be produced singly and on the other hand they
can decay into the Standard Model particles. The single production is
possible due to the mixing between the R-odd and R-even particles:
neutralinos mix with neutrinos, charginos with charged leptons, and
sneutrinos with Higgs bosons. The single production mechanisms provide
probes for heavier supersymmetric particles than one can explore when
R-parity is conserved in which case the production can only occur in
pairs. 

The phenomenology of spontaneously broken R-parity models at the
Z-peak at LEP has been previously considered in
\cite{majoron,BBHH}. In this paper we shall investigate the production
mechanisms of single neutralinos and charginos at LEP and a linear
collider in the framework of SLRM.

In Section 2 we will investigate the constraints for the mixing of the
Standard Model leptons with other constituents of neutralinos and
charginos obtained from the negative searches of single neutralino or
chargino production in LEP 1 measurements. We will not restrict
ourselves to the minimally extended MSSM, as was done e.g. in
\cite{majoron,BBHH}, but use a more general parameterization.  In
Section 3 we will first describe the basic ingredients of the
supersymmetric left-right model and then we will investigate in this
model the production and detection of neutralinos and charginos at
LEP2 and at an $e^+e^-$ linear collider.  A summary with conclusions
is presented in Section 4.

\section{Constraints from LEP 1 measurements}

In many supersymmetric models the lightest neutral supersymmetric
particle is a neutralino and the lightest charged one is a
chargino. This kind of models are of great interest from the point of
view of spontaneus R-parity breaking as large mixings between
neutrinos and the lightest neutralino and between charged leptons and
the lightest chargino might naturally occur there. Implications of
such mixings, if they exist, could in principle have been seen in LEP 1
experiments, and this can be used to set constraints on various model
parameters.  In this Section we will investigate those constraints.

We will assume that there exists a neutrino which is singlet under
Standard Model gauge group and which couples to one of the lepton
families of the Standard Model. The supersymmetric counterpart of the
singlet neutrino is assumed to acquire a nonzero VEV, driving the
spontaneous breaking of R-parity. The neutralino and chargino states
can be schematically written as superpositions of gauginos $\lambda_j
$, higgsinos $\tilde h_j$ and leptons $\nu_j$ and $ l_j$ as follows:

\bea \tilde \chi^0_i &=& a_{ij}^{(0)} \lambda^0_j + b_{ij}^{(0)}\tilde
h^0_j +c_{ij}^{(0)} \nu_j ,\nonumber \\ \tilde \chi^\pm_i &=&
a_{ij}^{(\pm)} \lambda^\pm_j + b_{ij}^{(\pm)} \tilde h^\pm_j +
c_{ij}^{(\pm)} l^\pm_j.  \eea

\noindent With typical values of the mass parameters in the
supersymmetric sector, i.e. of the order of the weak scale, one
assumes that the lightest neutralino, $\tco^0$, and the lightest
chargino, $\tco^+$, which are identified as the physical neutrino and
charged lepton, respectively, have interactions very similar to those
of the pure weak eigenstates of neutrino $\nu$ and charged lepton
$l^\pm$ in the SM.

Let us consider the production of the lightest "supersymmetric"
neutralino $\tc_{2}^{0}$ and chargino $\tc_{2}^{+}$, by assuming that
mass of the other states $(\tc_3^{0(+)},\tc_4^{0(+)},...)$ are beyond
the reach of LEP 1. The cross sections of $e^+e^- \rightarrow \tco^0
\tct^0$ $(\tco^\pm \tct^\mp)$ at $Z^0$ peak are suppressed by the
mixing of $\tc_{2}^{0}$ ($\tc_{2}^{+}$) with the neutrino
$\tc_{1}^{0}$ (the lepton $\tc_{1}^{+}$), but on the other hand
heavier neutralinos (charginos) can be produced than in pair
production with the same center of mass energy.  Of course, if the
mass of the neutralino or chargino is larger than $ M_Z/2$ the single
production is the only possibility to produce these particles at
$Z$-pole.

For the neutralino $\tct^{0}$ to be seen, it has to decay inside the
detector. If it is the lightest supersymmetric particle, it decays due
to the R-parity violating gauge interactions generated by its mixing
with the neutrino $\nu\simeq\tco^0 $.  Similarly, the decay of the
chargino, if it is the lightest supersymmetric particle, occurs as a
result of its mixing with a charged lepton $l^+\simeq\tc_1^+$. The
partial mean length of flight path of the neutralino and chargino in
the CM frame of $e^+e^-$ collison, taking only the decay via a virtual
$Z$ boson into account, is given by

\be
 L= \frac{s-m_{\tct}^2} {\eta^2 m_{\tct}^6 \sqrt{s}} \mbox{ } 6.7 \times 10^{-4}
\mbox{ } {\rm GeV}^5\cdot{\rm m}
\mbox{ } ,
\label{path}
\ee
\noindent
 where $m_{\tct}$ is the mass of the neutralino or chargino, $\sqrt{s}
$ is the CM energy of the collision, and the masses of neutrinos and
charged leptons are neglected.  The parameter $\eta$ is a suppression
factor which takes into account the fact that the $\tc_1\tc_2Z$
coupling is not of the full Standard Model strength. It is a measure
of the mixing between R-even leptons and R-odd supersymmetric
particles induced by the spontaneous R-parity breaking.

For typical values of the parameters the decay is very fast. The
production limit of a $\tco^0 \tct^0$ pair for the luminosity attained
at LEP 1 corresponds to the value $\eta=10^{-3}$ (see Fig.2). For a
neutralino with mass $m_{\tct^0}=45 $ GeV this would indicate the
decay length of $L=6\ \mu$m, that is, of the order of the sensitivity
of the LEP vertex detectors, for a heavier neutralino the decay length
is even shorter. The same is valid for the chargino decays. Hence,
whenever the neutralino or chargino coupling is large enough for the
single production at LEP, the produced particle also decays
immediately in the detector making its detection and identification
possible.

In Fig. 1 the R-parity breaking decay modes of the neutralino $\tct^0$
(assuming $m_{\tct^0}<m_{\tct^\pm}$) and of the chargino $\tct^\pm$
(assuming $m_{\tct^\pm}<m_{\tct^0}$) are depicted. It is assumed that
the lightest mass eigenstates $\tco^{0,\pm}$ are so close to the weak
interaction eigenstates $\nu,\,l^\pm$ that their other components can
be neglected in the first approximation. In this approximation the
three body decays of $\tct^{0,\pm}$ through virtual squark exchanges
($m_{\tilde q}>m_{\tct^{0,\pm}}$) can be ignored. Similarly the decays
into physical neutral or charged Higgses due to their slepton
components might be possible, but they are suppressed in general by
small mixings. Decays involving other components of the Higgs bosons
are proportional to the Yukawa type couplings, making the branching
ratios of these channels also generally unsignificant (but not always,
as we will see in next Section in the context of SLRM).  The
kinematically favoured decay of $\tct^{\pm}$ to $\tco^{\pm}$ is not
possible via $\gamma $ exchange, since the chargino states are
orthogonal and $\gamma $ couples similarly to all components of
$\tct^{\pm}$. Thus we are left with $\tct^{0,\pm}$ decay mode via
gauge boson $Z$ or $W$ exchange.

\input{epsf.tex}
\begin{figure}[t]
\leavevmode
\begin{center}
\epsfxsize=15.truecm\epsfysize=7.5truecm\epsffile{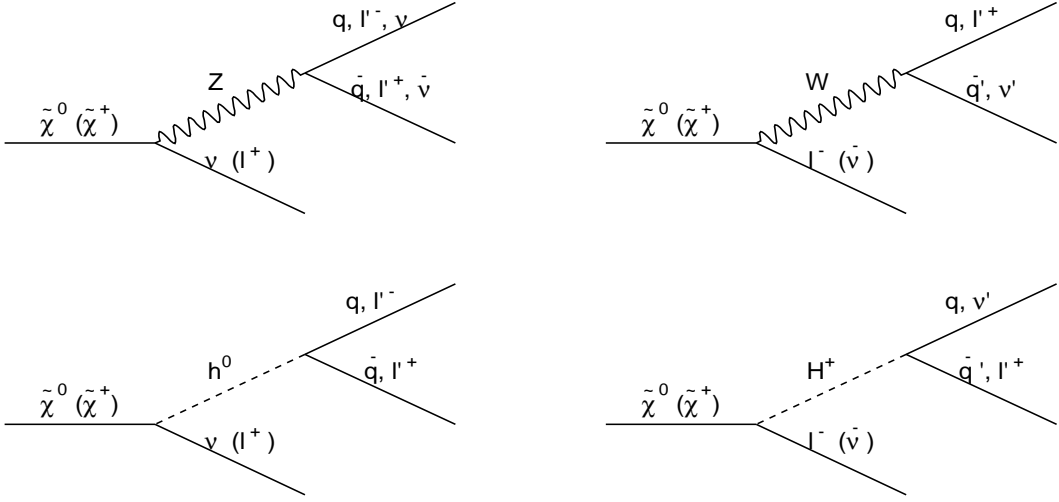}
\end{center}
\caption{\label{neutdec} The R-parity breaking decay modes of
neutralinos (charginos). The lightest neutralinos and charginos,
$\tco^{0,\pm}$, have been denoted by $\nu, \, l^\pm $. }
\end{figure}

The dominant final states resulting from the single production of the neutralino
$\tct^0$ in $e^+e^-$ collisions at $Z^0$ pole are thus the following:

\bea e^+e^- \rightarrow \bar\nu_l \tct^0 &\rightarrow&
\bar\nu_l\nu_{l'}  l^- l'^+ , \\ &\rightarrow& \bar\nu_l l^- q q'  ,\label{r02}\\
&\rightarrow&  \bar\nu_l\nu_l l'^+ l'^-  ,\\ &\rightarrow& \bar\nu_l\nu_{l} q\bar
q , \label{r04}\\ &\rightarrow& \bar\nu_l\nu_{l}\bar\nu_{l'}\nu_{l'} . 
\eea

\noindent Here $ \nu_l$ denotes the lightest neutralino $\tco^0$.  For
detection, the most favourable channel is (\ref{r02}) where the
visible invariant mass equals the mass of the decaying neutralino. The
background from the $\tau $ pair production can be effectively
eliminated by requiring that the invariant mass of the two jets is
larger than the $\tau $ mass.

In the case of a single $\tct^+$ production the dominant final states are

\bea
 e^+e^- \rightarrow \tct^+ l^- &\rightarrow& \nu\bar\nu l^+ l^- , \\
&\rightarrow& q\bar q'\bar\nu l^-  ,\label{r2}\\ &\rightarrow& \nu ' l'^+\bar\nu
l^-  ,\\ &\rightarrow& l'^+ l'^- l^+ l^- , \label{r4}\\ &\rightarrow&  q\bar q
l^+ l^-,  \label{r5}
\eea

\noindent where $l^-$ denotes ${\tilde\chi}_1^-$. For the first three
reactions, where part of the energy is invisible, there is a Standard
Model background from the $\tau $ pair production. In the case of the
reaction (\ref{r2}) this can be eliminated by a suitable cut in the
invariant mass of the hadronic jets. In the case of the (\ref{r4}) and
(\ref{r5}) reactions, all the energy is visible, allowing for a direct
chargino reconstruction.

The negative search of singly produced neutralinos and charginos at
LEP \cite{LEPdata} leads to constraints on the mass and couplings of
these particles. Considering the R-parity breaking production through
the dominant $Z$-exchange processes only, we present in Fig. 2 the
upper limit for the parameter $\eta$ introduced in Eq. (\ref{path}) as
a function of neutralino or chargino mass based on the total of 800
pb$^{-1}$ of data gathered by LEP at $Z$ pole \cite{LEPlum}. Taking
the detection limit of ten events and assuming a 100\% detection
efficiency, we find an upper limit of $\eta\lsim 0.01$, valid almost
up to the kinematical treshold.  

This simple analysis gives us a feeling of the sensitivity of the
present collider experiments on the effects of spontaneously broken
R-parity.  We will now move to the main part of our study. In the next
section we will analyse the phenomenology of spontaneously broken
R-parity at LEP2 and at a high-energy linear collider.  We shall
concentrate in our study on the supersymmetric $SU(2)_L \times SU(2)_R
\times U(1)_{B-L}$ model. As was pointed out in Introduction, in this
model R-parity is necessarily broken spontaneously \cite{KM,HM}.

\input{epsf.tex}
\begin{figure}[t]
\leavevmode
\begin{center}
\epsfxsize=15.truecm\epsfysize=12.5truecm\epsffile{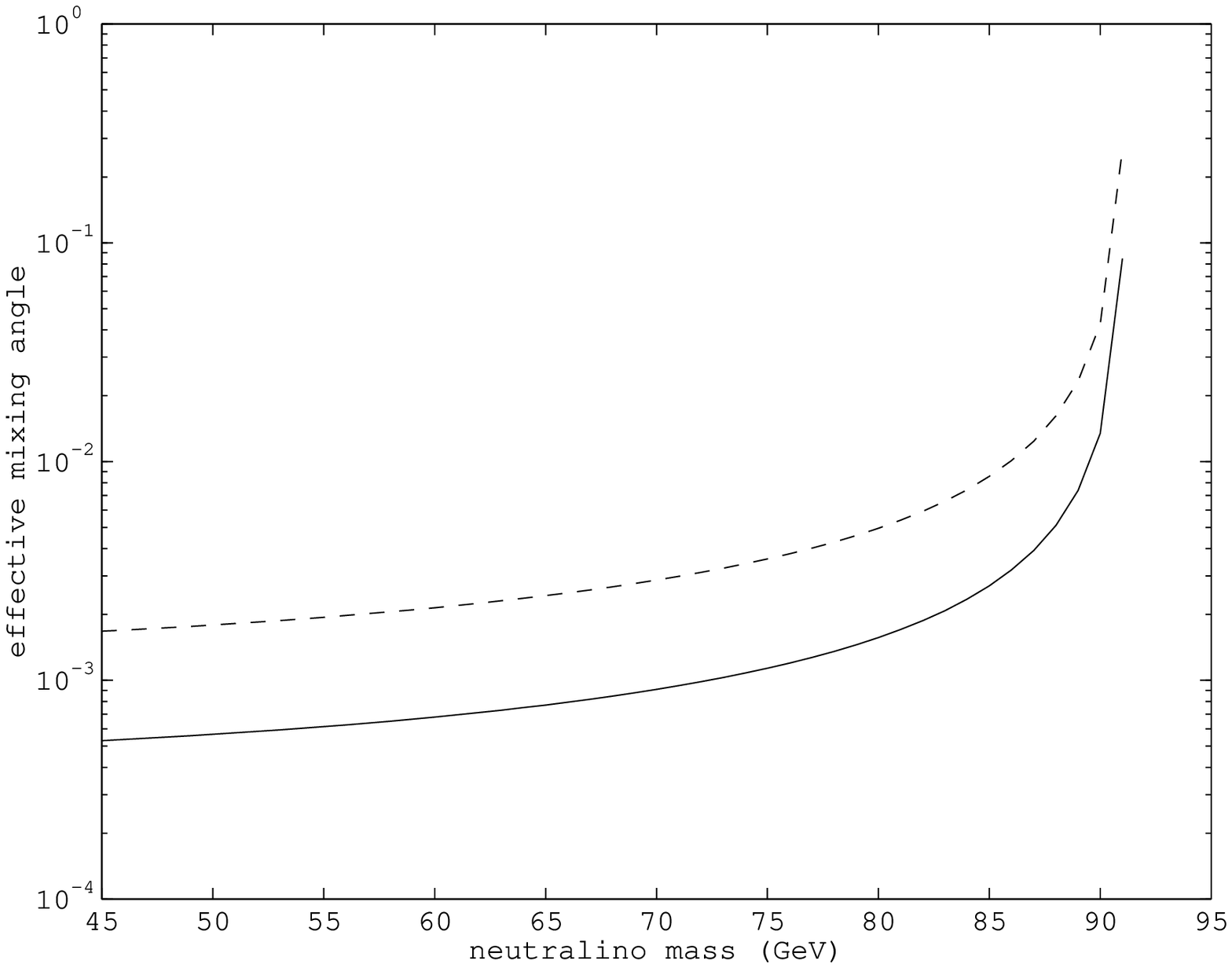}
\end{center}
\caption{\label{limit} The mixing between neutralino and neutrino (chargino and
lepton) for production of one neutralino (chargino) as a function of the
neutralino (chargino) mass (solid line). The dashed line corresponds to the
production of ten neutralinos (charginos)}
\end{figure}

\section{R-parity breaking effects at LEP2 and linear collider in SLRM}

\subsection{The model}

The framework of our analysis in the following is the SLRM defined by
the superpotential

\bea W & = & h_ {\phi }^Q Q^{T}i\tau_2 \phi Q^c + h_{ \varphi }^Q
Q^{T} i\tau_2 \varphi Q^c \nonumber \\ &&+h_ {\phi }^L L^{T}i\tau_2
\phi L^c +h_{ \varphi }^L L^{T} i\tau_2 \varphi L^c +h_{\DE} L^{cT}
i\tau_2 { \DE } L^c \nonumber\\ && + \mu_1 {\rm Tr} (i\tau_2 \phi^T
i\tau_2 \varphi ) +\mu_2 \Tr ( \DE \D ).\label{pot} \eea
 
\noindent 
Here $ Q$ denotes the left-handed quark superfield doublet, $
Q^c$ denotes the conjugate of the right-handed quark superfield
doublet and $L$ and $L^c$ are corresponding lepton
superfield doublets. The $B-L$ quantum numbers of these doublets are
+1/3, $-$1/3, $-$1 and +1, respectively. The triplet and the bidoublet
Higgs superfields are given by \bea
\label{higgses}
 \DE =\matr{cc} \DE^-/\sqrt{2} &  \DE^{0}\\
  \DE^{--}&- \DE^{-}/\sqrt{2} \ematr, 
\,\,\,\,\,\,\,\,\,\,\,\, && \delta =\matr{cc} \delta^{+}/\sqrt{2}&  \delta^{++}\\
  \delta^{0} &- \delta^{+}/\sqrt{2} \ematr  
\,\,\,\,\,\,\,\,\,\,\,\, \nonumber\\ &&\nonumber\\
\sim ({\bf 1,3,}-2), {\hspace*{2.truecm}}\,\,\,\,\,\,\,\,\,\,\,\,\,\,\,\,\,\,
\,\,\,\,\,\, && \sim ({\bf 1,3,}2),\nonumber\\ &&\nonumber\\
  \phi =\matr{cc} \phi_1^0&
 \phi_1^+\\ \phi_2^-& \phi_2^0 
\ematr
 \sim ({\bf 2,2,}0),
\,\,\,\,\,\,\,\,\,\,\,\, && \varphi =\matr{cc} \varphi_1^0&
 \varphi_1^+\\ \varphi_2^-& \varphi_2^0 
\ematr
  \sim ({\bf 2,2,}0).
\eea

In the level of superfields the R-parity conservation is equivalent to the
conservation of the multiplicative "matter parity"
\cite{Martin}
\be 
(\mbox{matter parity})= (-1)^{3(B-L)}.  
\ee 
Since $B-L$ is a generator of a gauge symmetry in the \ssu \\ model, the
Lagrangian of SLRM conserves R-parity automatically.  The spontaneous
breaking of the gauged U(1)$_{B-L}$ does not necessarily indicate
breaking of R-parity, but it was shown in \cite{KM} on general grounds
and in \cite{HM} explicitly in the model considered here that in order
to break the left-right symmetry in a physically consistent way also
R-parity must be broken.
 
The lowest state of the SLRM defined by the superpotential
$W$ in (\ref{pot}) corresponds to the following set of vacuum expectation values:

\be
\lan \phi_1^0\ran=\kappa_1,\,\,\lan \varphi_2^0 \ran =
\kappa_2,\,\,
\lan \Delta^0\ran = v_{\Delta},\,\, \lan \delta^0\ran =v_{\delta},\,\,
\lan \tilde\nu^c\ran =\sigma_R .
\label{vevs}
\ee

\noindent 
We have assumed here for simplicity that only one of the partners of
the right-handed neutrinos, $\tilde \nu_R$, gets a VEV. The neutrino
states of the corresponding family will then mix with neutral
higgsinos and the fermionic partners of the neutral gauge bosons,
$\lambda^0_L $, $\lambda^0_R$, and $\lambda^0_{B-L}$, to form the
physical neutralino states.

The mass Lagrangian of the neutralino sector can be written in the
form

\be {\cal{L}}=-\frac 12 \psi^T Y\psi +\mbox{h.c.},
\ee

\noindent where $\psi^T= (-i \lambda^0_L, -i \lambda^0_R , -i
\lambda^0_{B-L} , \tilde{\Phi}^0_1 , \tilde{\Phi}^0_2 ,
\tilde{\varphi}^0_1, \tilde{\varphi}^0_2 , \tilde{\Delta}^0
,\tilde{\delta}^0 , \nu , \nu^c)$ and the symmetric $11\times 11$
matrix $Y$ is given in Table 1.

\begin{table}[t]
\bea
\left(
\begin{array}{ccccccccccc} m_L & 0 & 0 & \frac{g_L
\kappa_1}{\sqrt{2}} & 0 & 0 & 
\frac{-g_L \kappa_2}{\sqrt{2}} & 0 & 0 & 0 & 0\\ & m_R  & 0 &
\frac{-g_R \kappa_1}{\sqrt{2}} & 0 & 0 & 
\frac{g_R \kappa_2}{\sqrt{2}} &  g_R v_\Delta \sqrt{2} & -g_R v_\delta \sqrt{2} &
0 &  -\frac {g_R\sigma_R}{\sqrt{2}}\\ && m_{V} & 0 & 0 & 0 & 0 & -g_{V} v_\Delta
\sqrt{2} & g_{V} v_\delta \sqrt{2} & 0 & \frac {g_V\sigma_R}{\sqrt{2}}\\ &&& 0 &
0 & 0 & -\mu_1 & 0 & 0 & 0 & 0 \\ &&&& 0 & -\mu_1 & 0 & 0 & 0 & h_{\Phi }^L
\sigma_R & 0 \\ &&&&& 0 & 0 & 0 & 0 & 0 & 0\\ &&&&&& 0 & 0 & 0 & h_{\varphi }^L
\sigma_R & 0
\\ &&&&&&& 0 & \mu_2 & 0 & -2 h_\Delta \sigma_R \\ &&&&&&&& 0 & 0 & 0 \\
&&&&&&&&& 0 & h_{\varphi }^L \kappa_2  \\ &&&&&&&&&& -2 h_\Delta v_\Delta 
\end{array}
\right) \nonumber
\eea
\caption{The upper triangle of the symmetric neutralino mass matrix $Y$. We use
the short-hand notation $V=B-L$.}
\end{table}

In addition to the mass terms that follow from eqs. (\ref{pot}) and
(\ref{vevs}), we have introduced in the matrix $Y$ diagonal soft
gaugino mass terms $m_L$, $m_R$ and $m_V$ for $\lambda^0_L$,
$\lambda^0_R$ and $\lambda^0_{B-L}$, respectively.

The lightest of the neutralinos corresponds to the left-handed
neutrino. As usual the masses of the other eigenstates
depend on the relative magnitudes of gaugino soft masses, parameters
$\mu_i$ and the VEVs (\ref{vevs}).

The mass Lagrangian of the chargino sector is given by

\begin{equation}
 {\cal{L}}=-\frac{1}{2} 
 (\varepsilon^{- T} \eta^{+ T}) 
\left( \begin{array}{c c} 0 & X \\ X^T & 0
\end{array} \right)
\left(
\begin{array}{l}
\varepsilon^- \\
\eta^+
\end{array}
\right) + \mbox{h.c.}
\end{equation}

\noindent where $\eta^{+T}=(-i \lambda_L^+,-i
\lambda_R^+,\tilde{\Phi}_1^+, \tilde{\varphi}_1^+,\tilde{\delta}^+,l^c
)$, and $\varepsilon^{-T}= (-i \lambda_L^-,-i \lambda_R^-,
\tilde{\Phi}_2^-, \tilde{\varphi}_2^-,\tilde{\Delta}^-, l)$. The
$6\times 6$ matrix $X$ is given in Table 2.

\begin{table}[t]
\bea
\left(
\begin{array}{ccccccc} m_L & 0 & 0 & g_L \kappa_2 & 0 & 0  \\ 0 & m_R & -g_R
\kappa_1 & 0 & g_R v_\delta \sqrt{2} & g_R\sigma_R \\ g_L \kappa_1 & 0
& 0 & \mu_1 & 0 & 0 \\ 0 & -g_R \kappa_2 & \mu_1 & 0 & 0 & 0 \\ 0 &
-g_R v_\Delta \sqrt{2} & 0 & 0 & \mu_2 & -h_\Delta \sigma_R \sqrt{2}
\\ 0 & 0 & -h_{\Phi }^L \sigma_R & -h_{\varphi }^L \sigma_R & 0 &
-h_{\Phi }^L \kappa_1
\end{array}
\right)\nonumber
\eea
\caption{The chargino mass matrix $X$.}
\end{table}

In models with R-parity conservation, the lightest neutralino is
usually the lightest supersymmetric particle, LSP. In our case where
$R$-parity is spontaneously broken, it may well happen that the
lightest chargino $\tct^+$ is lighter than $\tct^0$.

As a result of the spontaneous R-parity breaking, the neutralinos and
charginos corresponding to the SM leptons, being superpositions of
gauginos, higgsinos and leptons, have in general couplings different
from the couplings of the SM leptons. The precision measurement of
those couplings made, e.g., at the Z-boson pole at CERN should be
taken into account when constructing realistic models. Apart from the
precision collider measurements also some rare processes lead to
meaningful constraints. The most stringent bound is obtained from the
non-observation of the exotic muon decay $\mu \rightarrow 3e$ whose
branching ratio obeys Br$(\mu \rightarrow 3e)<10^{-12}$ \cite{PDG}. If
$\langle \tilde{\nu}_e \rangle$ and $\langle \tilde{\nu}_{\mu}\rangle$
are both large this constraint is not satisfied in SLRM.

It is interesting to note in this connection that in SLRM the
components which usually form the dominant part of the mass
eigenstates of charginos and neutralinos corresponding to the SM
leptons have approximately the SM couplings with $SU(2)_L$ gauge
bosons due to the symmetry breaking structure. This is because the
states that diagonalize the Lagrangian after the $SU(2)_R$ breaking
have definite $SU(2)_L \times U(1)_Y$ quantum numbers and
couplings. The subsequent breakdown to $U(1)_{em}$ cause only small
deviations, characterized by $(m_{W_L} \mbox{ or } m_\tau)/m_{W_R}$ or
$m_\tau/m_{susy}$, where $m_{susy}$ is a general susy scale, to the
composition of these states and their couplings.

\subsection{Production of neutralinos and charginos at LEP 2 and  at a linear
collider}

Predictions of the SLRM depend on the values of the soft gaugino
masses $m_L$, $m_R$ and $m_V$ which are in principle free
parameters. We will consider in the following two representative
choices of these parameters.  In one case the soft gaugino masses are
all of the order of 100 GeV (Model 1) and in the other one of the
order of 1 TeV (Model 2). The masses and compositions of the physical
particles relevant for the decays of neutralinos and charginos are
listed for both models in Table 3. We have assumed for simplicity that
only the tau sneutrino $\tilde{\nu}_\tau$ achieves a non-zero VEV.

\begin{table}
\begin{tabular}{c|r|c} Model 1 && \\ particle & mass [GeV] & composition \\
\hline
$ {Z_2}$& 1654 &   \\
$ {W_2}$&1029 &  \\
$ {\tco^0}\sim \nu_\tau $ & 0.02  & $(-0.01i \lambda_L^0+0.01 i
\lambda_R^0 -0.05
\tilde\phi_1^0-\nu)$\\
$ {\tct^0}$ & 93  & $(0.98 i \lambda_L^0-0.11 i \lambda_R^0-0.17 i
\lambda_{B-L}^0 $\\ &&$+0.03
\tilde\phi_1^0-0.02\tilde\varphi_2^0-0.01\nu)$\\
$ {\tco^+\sim \tau^+ }$ & 1.7  & $\left( \begin{array}{c} 0.79
i\lambda_R^+-0.03\tilde\phi_1^+ -0.3\tilde\delta^++0.54l^c\\
0.01i\overline{\lambda_L^-}-0.01i\overline{\lambda_R^-}
-0.03\overline{\tilde\phi_2^-}-0.05\overline{\tilde\varphi_2^-}
+0.01\overline{\tilde\Delta^-}-\overline{l}
\end{array}\right)$\\
${\tct^+}$ & 94  & $\left( \begin{array}{c}  i\lambda_L^++0.02\tilde\varphi_1^+\\
-i\overline{\lambda_L^-} +0.04\overline{\tilde\phi_2^-}-0.01\overline{l}
\end{array}\right)$\\
${H_1^0}$ & 53  &$(-0.03\tilde\nu^c-0.47\phi_1^0-0.88\varphi_2^0-0.02\Delta^0
-0.04\delta^0$)\\
\hline Model 2&& \\ particle & mass [GeV] & composition \\
\hline
$ {Z_2}$& 962 & \\
$ {W_2}$& 742 &\\
$ {\tco^0}\sim \nu_\tau $ &  $0.001 $ & $(0.04
\tilde\varphi_1^0 +\nu )$ \\
$ {\tct^0}$ & 
$333$  & 
$ ( 0.38 i\lambda_R^0 
-0.25 i\lambda_V^0 
+0.22 \tilde\Delta^0-0.82 \tilde\delta^0 
-0.29 \nu^c $
\\
$ {\tco^+}\sim\tau^+$ & 1.7  & $\left( \begin{array}{c} -0.6i{\lambda_R^+}
-0.68\tilde\delta^+-0.42l^c\\ -0.04\overline{\tilde\varphi_2^-} -\overline{l}
\end{array}\right)$\\
${\tct^+}$ & 772  & $\left( \begin{array}{c} -0.61i{\lambda_R^+}
+0.73\tilde\delta^+-0.3l^c\\ 0.71 i\overline{\lambda_R^-}
+0.71\overline{\tilde\Delta^-}\end{array}\right)$\\
${\tilde H^{++}}$ & 500  &$\left( \begin{array}{c} 
\tilde\delta^{++}\\
\overline{\tilde\Delta^{--}}\end{array}\right)$\\
${H_1^0}$  & 108  &
$(0.02\tilde\nu^c+0.02\phi_1^0+\varphi_2^0-0.01\Delta^0 +0.01\delta^0 $) \\
$ {H^{++}}$  & 334  & ($0.91\Delta^{++}+0.42\delta^{++}$) \\
\end{tabular}
\caption{Masses and compositions of particles with 
$m\lsim m_{\tilde\chi_2^{\pm,0}}$ and masses of the heavy gauge bosons in Models
1 and 2. In Model 1, the soft gaugino masses
$M_i\simeq 100$ GeV and
$\kappa_2/\kappa_1=1.9$ and $\sigma_R=1.7$ TeV. In Model 2, the soft gaugino masses
$M_i\simeq 1$ TeV and 
$\kappa_2/\kappa_1 =50$ and $\sigma_R=340$ GeV. Yukawa coupling $h_\Delta =0.6$,
$v_\Delta/v_\delta \simeq 1.1$ and the soft scalar masses are 
${\cal{O}}(1$ TeV) in both models. }
\end{table}

Let us consider the two models separately.

\begin{figure}[t]
\leavevmode
\begin{center}
\epsfxsize=15.truecm\epsfysize=4.7truecm\epsffile{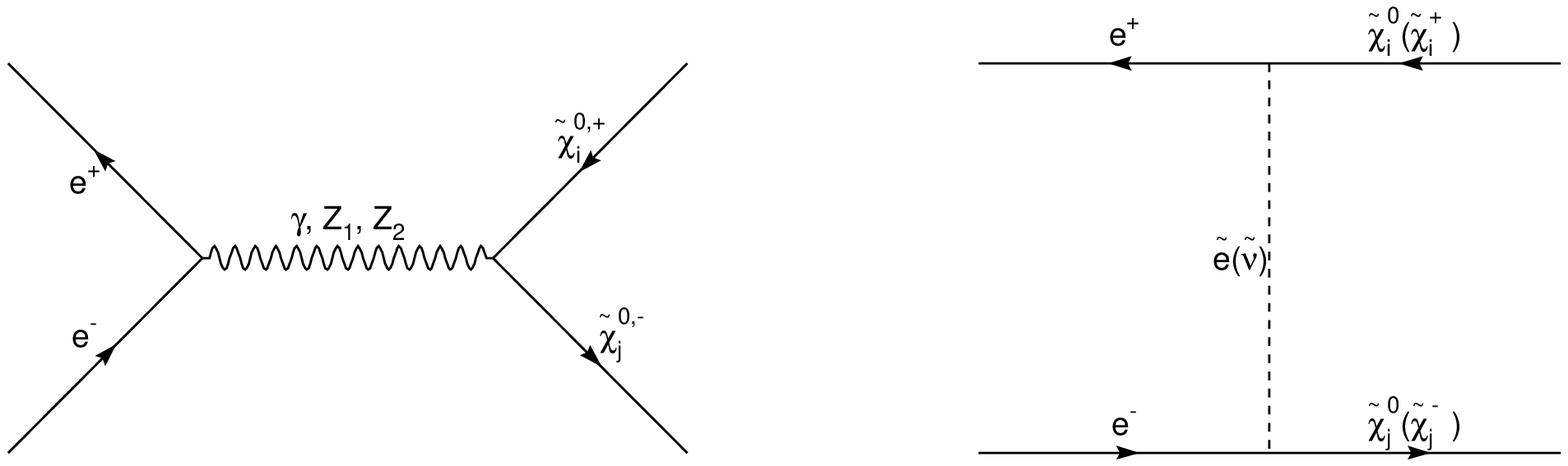}
\end{center}
\caption{\label{prod} The Feynman graphs for single and pair
production of neutralinos and charginos
in SLRM. }
\end{figure}

\begin{figure}[t]
\leavevmode
\begin{center}
\epsfxsize=15.truecm\epsfysize=15.truecm\epsffile{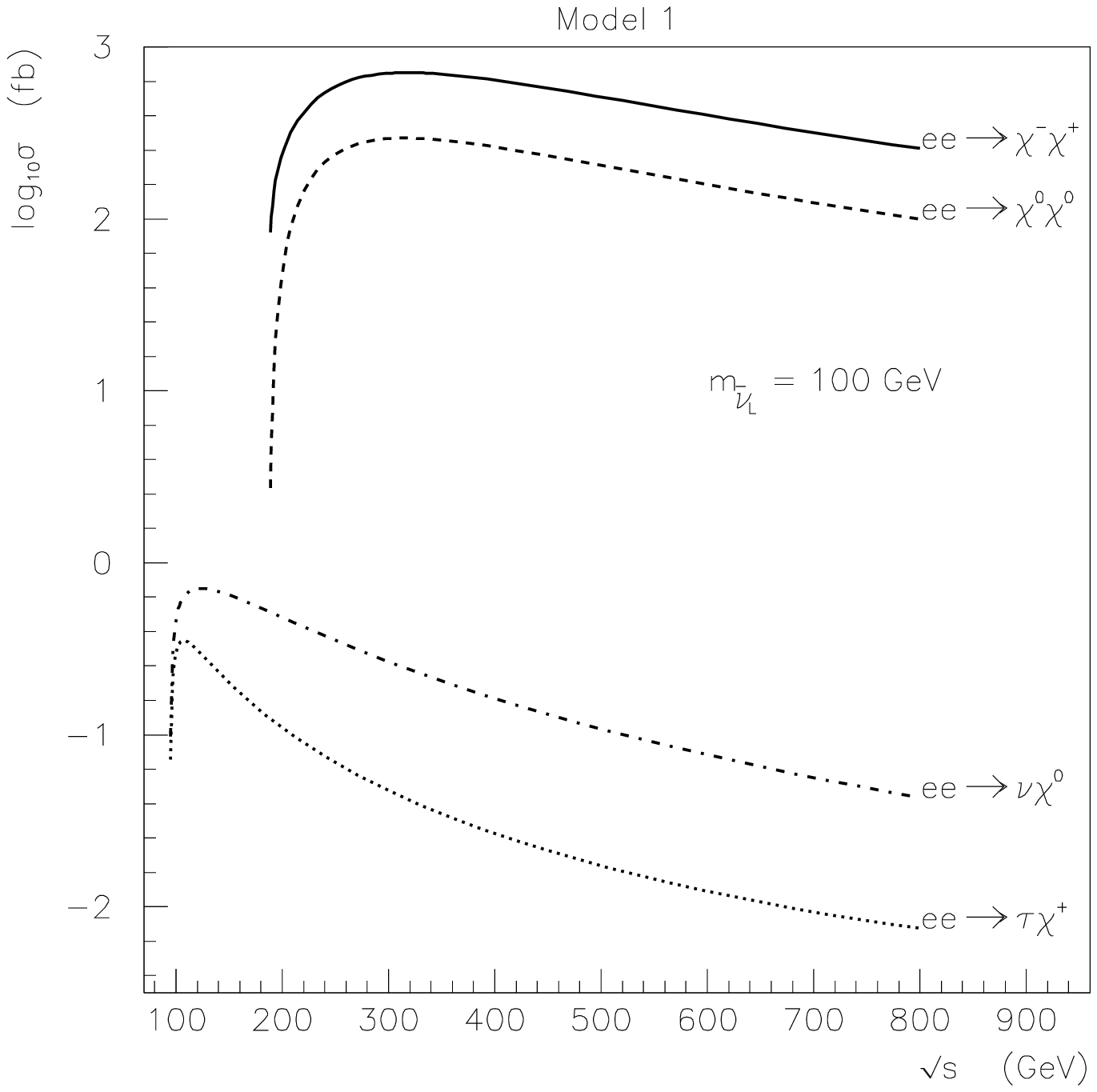}
\end{center}
\caption{\label{M1csa}  The production cross section of single or pair produced 
neutralinos and charginos  as a function of center of mass energy in
the Model 1 with $m_{\tilde e_L}=m_{\tilde \nu_{eL}}=100$
GeV. 
}
\end{figure}

\begin{figure}[t]
\leavevmode
\begin{center}
\epsfxsize=15truecm\epsfysize=15truecm\epsffile{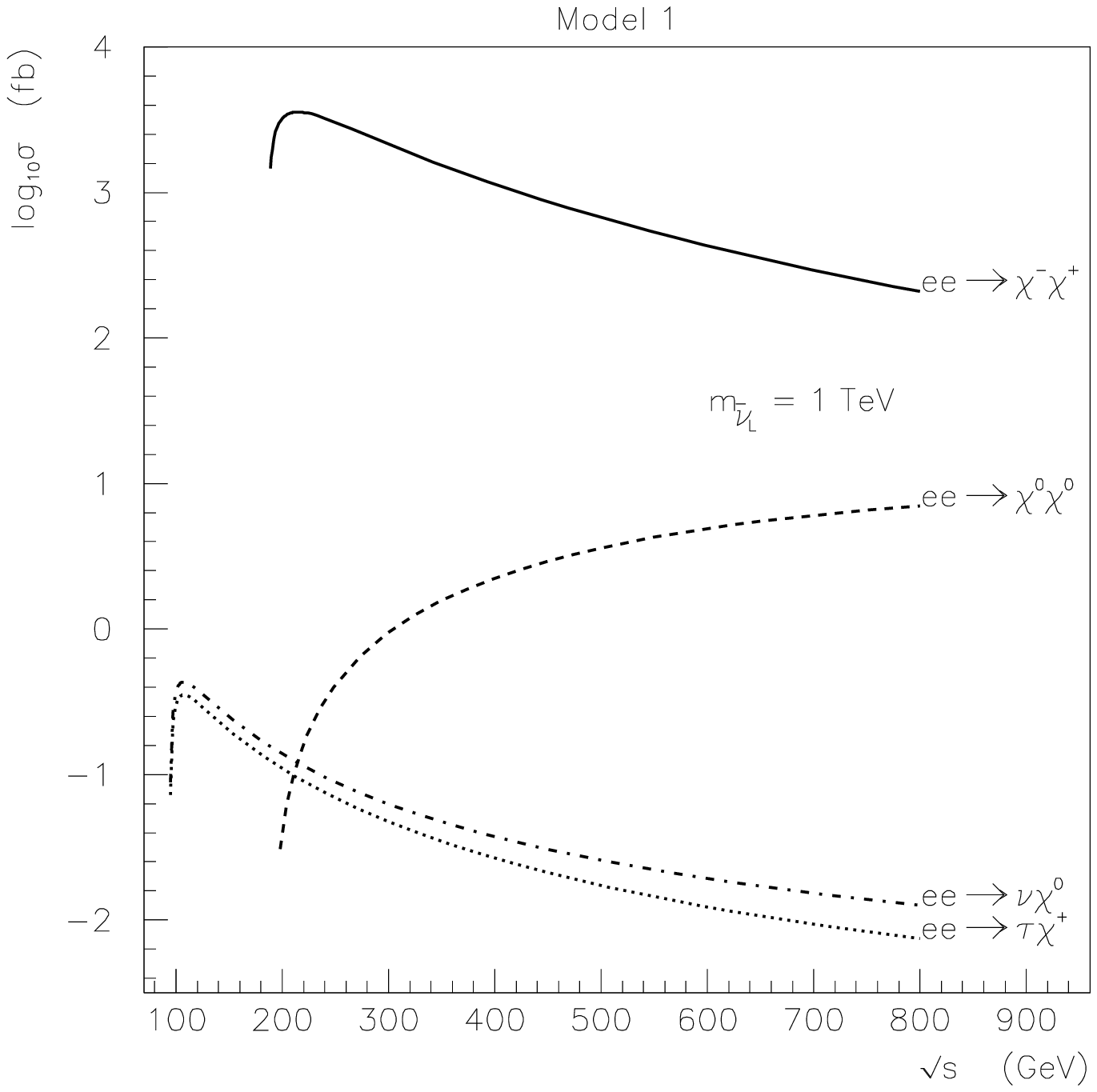}
\end{center}
\caption{\label{M1csb}  The production cross section of single or pair produced 
neutralinos and charginos  as a function of center of mass energy in
the Model 1 with $m_{\tilde e_L}=m_{\tilde \nu_{eL}}=1$
TeV. 
}
\end{figure}

{\bf Model 1}.  In Model 1 the lightest supersymmetric chargino
$\tct^\pm$ and the neutralino $\tct^0$ are almost degenerate: their
masses are $m_{\tct^\pm}\simeq 94$ GeV and $m_{\tct^0}\simeq 93$ GeV,
and they are both almost pure gaugino $\lambda_{L}$ states.  The
possible production amplitudes are depicted in Figure \ref{prod}. In
Figs. \ref{M1csa} and \ref{M1csb} we have plotted the cross section
for the single production via the reactions $e^+e^-\to \tct^+\tau^-$
and $e^+e^-\to \tct^0\nu_{\tau}$ and for the pair production via the
reactions $e^+e^-\to \tct^+\tct^-$ and $e^+e^-\to \tct^0\tct^0$ as a
function of the center of mass energy $\sqrt{s}$.
The single production occurs via $Z_L$ exchange in s-channel and via
sneutrino or selectron exchange in t-channel.  For the chargino pair
production there is an additional contribution coming from the
s-channel photon exchange.

At LEP 1 the collision energy is not large enough for the production
of $\tct^{\pm}$ or $\tct^0$, but at LEP 1.5 the single production is
possible in principle. From Figs. \ref{M1csa} and \ref{M1csb} one can
see, however, that while the single production is kinematically
allowed at the LEP 1.5 energies of $\sqrt{s}=130 - 140$ GeV, the cross
section is far too small for a signal with the collected data of about
5 pb$^{-1}$. Even for LEP 2 with $\sqrt{s}=160 - 192$ GeV and the
planned integrated total luminosity of 500 pb$^{-1}$, the cross
section is less than 1 fb and no events would be produced during the
whole running time of LEP 2.

Contrary to the single production, the pair production is not
suppressed by mixing angles and consequently the cross section of pair
production may exceed that of the single sparticle production by
several orders of magnitude.  As seen from Figs.  \ref{M1csa} and \ref{M1csb}, the
chargino pair production is a particularly promising process. The
cross section depends on the mass of the electron sneutrino
$\tilde{\nu_e}$, and in Figs.  \ref{M1csa} and \ref{M1csb} we have plotted the cross
section for two representative values, $m_{\tilde\nu_e}=100$ GeV and
$m_{\tilde\nu_e}=1$ TeV, respectively. The destructive interference between the
s-channel and t-channel contributions reduces the cross section
considerably in the case of a light sneutrino, but it is negligible
for a heavy sneutrino case. At the center of mass energy of $\sqrt{s}=
192$ GeV the cross section of the pair production process $e^+e^-\to
\tct^+\tct^-$ in the Model 1 assuming $m_{\tilde{\nu_e}}= 1$ TeV is
2.2 pb, resulting in hundreds of signal events.

In Model 1 the neutralino decays mainly to the lightest chargino
$\tco\simeq \tau$ and $W_1$ with the decay width of
$\Gamma(\tct^0\rightarrow \tco^+W_1) \simeq 10$ keV. The chargino can
decay either to the $W_1$ and the lightest neutralino $\simeq
\nu_\tau$ for which $\Gamma (\tct^+\rightarrow \tco^0W_1) \simeq 20$
keV or to the lightest chargino and Higgs boson for which $\Gamma
(\tct^+\rightarrow \tco^+H_1^0) \simeq 10$ keV, followed by the decay
of Higgs to a pair of bottom quarks.  It is evident that the signal
for chargino pair production should be detectable at LEP 2 with center
of mass energy 192 GeV and integrated luminosity 200 pb$^{-1}$, as
well as at a linear collider with anticipated center of mass energies
of 350 to 1600 GeV and luminosities of 10 to 500 fb$^{-1}$
\cite{Wiik}, respectively, even if only the channels with all the
final state energy visible is used for the detection.

\begin{figure}[t]
\leavevmode
\begin{center}
\epsfxsize=15.truecm\epsfysize=15.truecm\epsffile{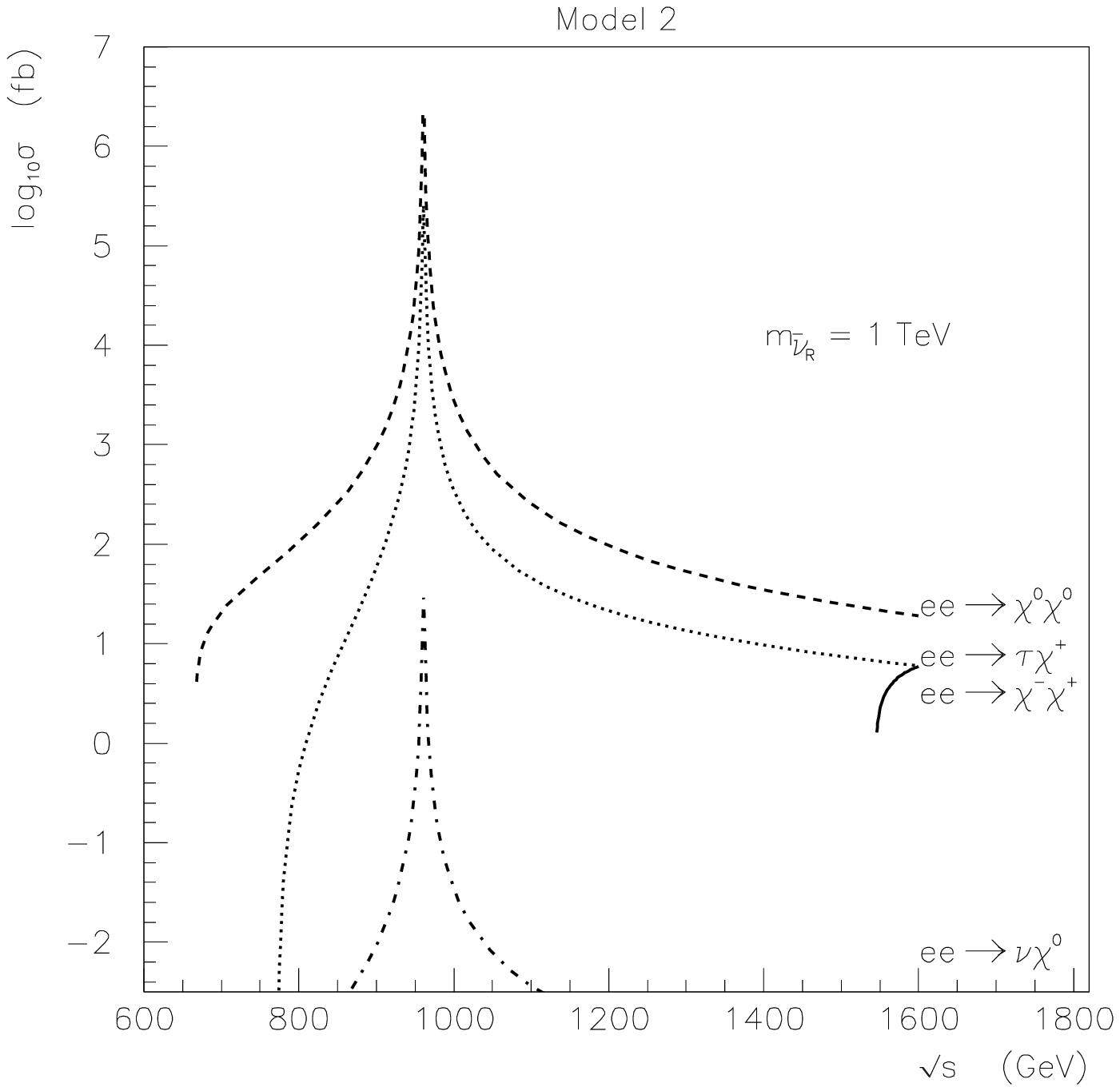}
\end{center}
\caption{\label{M2cs} The production cross section of single or pair
produced neutralinos and charginos as a function of center of mass
energy in Model 2 with $m_{\tilde e_R}=m_{\tilde \nu_{eR}}=1$ TeV. }
\end{figure}

{\bf Model 2.} In this model the soft gaugino masses are of the order
 of 1 TeV. The masses of the lightest supersymmetric chargino and
 neutralino $\tct^{\pm}$ and $\tct^0$ are 772 GeV and 333 GeV,
 respectively, that is, they are too heavy to be produced at LEP, even
 singly. At a linear collider with $\sqrt{s}$ up to 1.6 TeV they can
 be produced both in single or pair production processes.  The
 production cross sections are plotted in Fig. \ref{M2cs}. The mass of
 the heavy neutral weak boson $Z_2$ in this model is 962 GeV,
 resulting in a resonance peak in the cross section in the energy
 range relevant for the future linear collider. The dominant processes
 are now $e^+e^-\to \tct^0\tct^0$ and $e^+e^-\to \tct^+\tau$ whose
 cross sections are at the level of tens of fb's outside the resonance
 region. In Fig. \ref{M2cs} we have taken the width of the $Z_2$
 resonance to be equal to that of the ordinary $Z$ boson. The single
 production of the neutralino in this model is negligible, and for the
 most part of the interesting energy range the pair production of the
 chargino is kinematically excluded.

 The dominant decay channel of the neutralino $\tct^0$ in Model 2 is
 to the lightest neutralino and the lightest Higgs boson, for which
 the decay width is $\Gamma (\tct^0\rightarrow \tco^0 H_1^0) \simeq
 100$ keV. Branching ratios to gauge bosons are in this case more
 suppressed than in the Model 1, the width being $\Gamma
 (\tct^0\rightarrow \tco^0Z_1) \simeq \Gamma (\tct^0\rightarrow
 \tco^+W_1) \simeq 30$ keV. The chargino decays in Model 2
 predominantly to a doubly charged Higgs boson $H^{++}$, whose mass is
 334 GeV, and a charged lepton followed by the decay of $H^{++}$ to a
 same sign lepton pair. The width of this channel is $\Gamma( \tct^+
 \rightarrow \tco^- H^{++} ) \simeq 90$ MeV. These decays lead to a
 spectacular signature characteristic of SLRM. The kinematically most
 favoured decay mode of the doubly charged Higgs is $H^{++}\to ll'$.
 In the case $l=l'=e$ or $l=l'=\mu$ one will have from this decay an
 unambiguous signature of a same-sign lepton pair with no missing
 energy. In the case $l=l'=\tau$ the final state would include either
 two same-sign leptons ($e^-e^-$, $\mu^-\mu^-$ or $e^-\mu^-$) with
 missing energy or transverse pions plus missing energy.  The decay
 widths for the other channels are $\Gamma (\tct^+\rightarrow
 \tco^+Z_L) \simeq 3$ MeV, $\Gamma (\tct^+\rightarrow \tco^0W_1^+)
 \simeq 1$ MeV, and $\Gamma (\tct^+\rightarrow \tco^+ H^0_1) \simeq
 0.5$ MeV. If the right-handed slepton is light enough the chargino
 will decay via R-parity preserving processes into a left-handed
 slepton and a lepton.

Background for the conventional R-parity violating decay modes,
discussed in the case of LEP 1, increases substantially when the
center of mass energy is beyond the gauge boson pair production
threshold. In the case of separate lepton number violating decays
involving doubly charged Higgses there is no Standard Model background
for the process.

\section{Summary and conclusions}

We have studied the production and decay of neutralinos and
charginos at LEP 2 and at a linear collider in the case of spontaneously broken 
R-parity within the framework of supersymmetric \ssu model.
Characteristic of models with  R-parity non-conservation is that missing energy
is no longer a signature in all of the supersymmetric processes. The signal in
SLRM may be similar to the case of MSSM with R-parity breaking (Model 1), but for
an interesting part of the parameter space the signals typical for
the left-right model, namely the decay via a doubly charged
Higgs, is the dominant one (Model 2).

\vspace{2cm}
\large
\noindent Acknowledgements
\normalsize
\vspace{0.5cm}
 
The work has been supported by the  Academy of Finland.

\end{document}